\documentclass[12pt,preprint]{aastex}

\usepackage{color}

\begin{document}

\title{The Core Collapse Supernova Rate from the SDSS-II Supernova Survey}

\date{July 1, 2014}

\author{
Matt~Taylor\altaffilmark{1}, 
David~Cinabro\altaffilmark{1},
Ben Dilday\altaffilmark{2},
Lluis~Galbany\altaffilmark{3},\altaffilmark{4},
Ravi~R.~Gupta\altaffilmark{5},
R.~Kessler\altaffilmark{6},\altaffilmark{7},
John~Marriner\altaffilmark{8},
Robert~C.~Nichol\altaffilmark{9},
Michael~Richmond,\altaffilmark{10}, 
Donald~P.~Schneider\altaffilmark{11},\altaffilmark{12},
and,
Jesper~Sollerman\altaffilmark{13}
}

\altaffiltext{1}{
Department of Physics and Astronomy, Wayne State University, Detroit, MI 48202 USA}
\altaffiltext{2}{
North Idaho College, 1000 W. Garden Ave. Coeur d'Alene, ID 83814 USA}
\altaffiltext{3}{
Millennium Institute of Astrophysics, Universidad de Chile, Casilla 36-D, Santiago, Chile}
\altaffiltext{4}{
Departamento de Astronomia, Universidad de Chile, Casilla 36-D, Santiago, Chile}
\altaffiltext{5}{
Department of Physics and Astronomy, University of Pennsylvania, Philadelphia, PA 19104 USA}
\altaffiltext{6}{
Department of Astronomy and Astrophysics, University of Chicago, Chicago, IL 60637 USA}
\altaffiltext{7}{
Kavli Institute for Cosmological Physics, University of Chicago, Chicago, IL 60637 USA}
\altaffiltext{8}{
Center for Particle Astrophysics, Fermi National Accelerator Laboratory, P.O. Box 500, Batavia, IL 60510 USA}
\altaffiltext{9}{
Institute of Cosmology \& Gravitation, Dennis Sciama Building, University of Portsmouth, Portsmouth, PO1 2FX, UK}
\altaffiltext{10}{
School of Physics and Astronomy, Rochester Institute of Technology, Rochester, NY 14623 USA}
\altaffiltext{11}{
Department of Astronomy and Astrophysics, The Pennsylvania State University, University Park, PA 16802 USA}
\altaffiltext{12}{
Institute for Gravitation and the Cosmos, The Pennsylvania State University, University Park, PA 16802 USA}
\altaffiltext{13}{
The Oskar Klein Centre, Department of Astronomy, AlbaNova, SE-106 91 Stockholm, Sweden}

\email{cinabro@physics.wayne.edu}

\newpage

\begin{abstract}
We use the Sloan Digital Sky Survey II Supernova Survey (SDSS-II SNS) data to
measure the volumetric core collapse supernova (CCSN) rate in the
redshift range $(0.03<z<0.09)$.  Using a sample of 89 CCSN we find 
a volume-averaged rate of
$1.06 {\pm} 0.19 \times 10^{-4} \frac{(h/0.7)^3}{(\mathrm{yr} \mathrm{Mpc}^3)}$
at a mean redshift of $0.072 \pm 0.009$.
We measure the CCSN luminosity function from the data and
consider the implications on the star formation history.
\end{abstract}



\setcounter{footnote}{0}

\section{Introduction}
\label{sec:intro}

Supernovae of the observational Types Ib, Ic and all Type II are the result
of core collapse events in massive stars.  Because the interval
from massive star formation to CCSN is brief
on astronomical time scales, the CCSN distribution in space and time
traces the formation of massive stars, a process that is not currently
well understood.  Study of the stellar population in galaxies shows an
increasing rate of star formation with redshift up to $z \approx 0.5$
proportional to the phenomenological rate formula $(1+z)^\beta$
with $\beta$ having values in the range 2.5 to 3.9 \citep{hopkins2004,schiminovich,
lefoch,hopkins2006, rujopakarn, cucciati} 
depending on the redshift and wavelength ranges considered.
Comparisons between the star formation and CCSN rate densities show that the
redshift dependence of the two agree well up to redshift of about 1.0
with higher values of $\beta$ as in \citet{horiuchi}.

However, there is confusion on the absolute comparison between the two rates:
when the overall star formation rate density and initial mass function are used to 
predict the rate at which massive stars form, it {\it ought} to match the
rate at which CCSN  occur, given our current 
understanding of stellar evolution.
Horiuchi and collaborators in \citet{horiuchi} note 
an apparent discrepancy between the two absolute rates with the star formation
implying a much higher CCSN rate than is observed.  
This disagreement could be caused by a very large misunderstanding of stellar evolution,
a large change in the initial stellar mass function,
or large population of dim or extincted CCSN that evade detection as a function of
redshift that preserves the observed redshift dependence. 
Botticella and collaborators in \citet{botticella2012}
conclude that the star formation rate derived from H$\alpha$ is too 
small by a factor of 2, which would exacerbate further
the discrepancy reported in \citet{horiuchi}.  Recently Mathews
and collaborators in \citet{mathews} have looked at this and
conclude that there may no ``supernova rate problem'' at all.

The goal of this work is to measure the CCSN rate more accurately than was
previously possible, taking advantage of the large spatial volume sampled
by and the uniform observing conditions of the SDSS-II SNS described in \citet{frieman}, to
help resolve this confusion.  More properly we will measure the CCSN rate
density, but we often use the common short hand of ``rate'' for the rate density.
The SDSS-II SNS was designed to detect type Ia supernovae (SNIa) in the
so-called ``redshift desert", where previous supernova data were lacking.
Likewise, the survey's design allows our CCSN rate measurements to probe an
intermediate redshift range where no prior CCSN rate has been published.
Our CCSN data span the region $0.03 < z < 0.09$.  
Combined with other CCSN rate measurements, our measurement helps
anchor the CCSN rate desnity versus redshift curve.

The CCSN rate density is also valuable input to supernova cosmology studies.
Increasingly large samples of photometric supernova light curves
are being gathered with only a small fraction of the objects having
simultaneous spectral data for the identification of supernova types.
Precise knowledge of the CCSN rate will be needed to understand
and quantify the level of CCSN contamination in a sample of SNIa 
selected with only photometry.

The current generation of supernova surveys
have greatly increased the accuracy and time resolution
of supernova observations.  The SDSS-II SNS is part of this cohort, along with
the Supernova Legacy Survey (SNLS) and Southern intermediate redshift ESO
Supernova Search (STRESS).  SDSS-II SNS and SNLS were designed primarily
to measure SNIa candidates for cosmology, but incidentally detected
a large, well--characterized sample of CCSN.  STRESS, on the other
hand, was explicitly designed as a supernova rate survey for both
SNIa and CCSN, though it did not quite have the same commitment of
observational resources as SNLS and SDSS-II SNS.  The SNLS analysis
of \citet{bazin} and the STRESS analysis of 
\citet{botticella}
are today's state of the art CCSN rate density
measurements.  This work adds to the CCSN rate
results for the current generation of surveys.

Surveys in search of SNIa provide an opportunity to measure the
CCSN rate density, as the
observation and image processing pipeline designed to discover SNIa will
incidentally detect a nearly equal number of CCSN.  We have employed this
approach using data from the SDSS-II SNS, 
from which we extracted ${\sim}10000$ CCSN candidates, although
we use only a fraction of this sample to measure the CCSN rate at
redshift less than $0.1$.

In the next sections we describe the
SDSS-II SNS and the important ancillary data we obtained
from the SDSS-III BOSS project described in \citet{eisenstein,bossover}, 
how we characterize candidates, 
our model for the efficiency of detecting CCSN,
the selections that we make to count observed CCSN,
corrections we apply to that count, 
sources of uncertainty, 
the rate calculation,
discuss our result,
and conclude.

\section{The SDSS-II SNS and SDSS-III BOSS}
\label{sec:sndetect}

We only briefly describe the SDSS-II SNS here.  Full details are given
in \citet{frieman,sdss1,sdss2,sdss3,sdss4}.
\citet{sako2014} makes the entire data sample publically 
available and gives more details about its composition.
For the SDSS-II SNS, a 300 $\mathrm{deg}^2$ region of sky, 
designated the Equatorial stripe and also known as `Stripe 82', 
was targeted for imaging once every two days;
however, viewing conditions only allowed imaging of the entire stripe
once every four days, on average.
To detect supernovae within the search region, images from each night
were compared to a previously observed SDSS template
image of the same region of the sky to identify transient objects.
To reduce the number of candidates to consider, a catalog of known quasars, variable stars and
active galactic nuclei (AGNs) was used to exclude variable objects that
are known not to be supernovae.  We account later for these known variables by
reducing our observing area.  We hand scanned the images of the excluded AGNs,
and observed no additional variable objects within them. 
Also, most objects within the solar
system are rejected by software; their proper motion is so large that
their position shifts significantly in the few minutes between $g-$,
$r-$ and $i-$band exposures.  

Initially the observed variable objects were forwarded to a team of human scanners
within the collaboration.  Images of each object were visually inspected
by one or more
scanners, who registered their judgment on whether the object
might be a supernova.  Many transient signals are obviously not supernovae,
including fast moving objects, poor image subtractions,
and telescope artifacts.  Also, when a variable object was detected
in more than one year of observation, it was excluded from the sample,
as supernovae are extremely unlikely to be detected over such a long period of
time.  As the survey progressed, exclusion of these non-supernovae became
increasingly automated, so that a greater fraction of objects forwarded
to the scanners were subsequently identified as possible supernovae.  
A key change made after our first observing season is that software
observed variable objects were only forwarded to human scanners after two
observations, rather than one.  This reduced the number of human scanned
objects by nearly a factor of ten, but led to little change in the number
of supernova candidates, and no change for variable objects with peak
magnitude brighter than magnitude 21 in the $r-$band. 
The photometry for the supernova candidates is described in \citet{holtzman}.

As the SDSS-II SNS was designed to detect and measure SNIa,
it produced an excellent sample for measuring the rate density
of SNIa.  Analysis of the SDSS-II SNS data 
in \citet{dilday2008} and \citet{dilday2010}
identified a rate sample of over 500 SNIa
candidates, approximately half of which were spectroscopically confirmed.
The remainder were typed by a template fitting algorithm, shown in Monte
Carlo simulations to add only about 5\% uncertainty to the rate
measurement due to false positives.

For the CCSN rate measurement presented here,
we make use of methods developed in \citet{dilday2008}.  In general, CCSN
are much more diverse than SNIa, and therefore tools to identify CCSN
candidates from photometry alone are not as well developed as for SNIa.
However, we can make use of the current consensus view that SNIa and
CCSN together comprise nearly the entire set of observed supernovae besides
``peculiar'' SNIa and CCSN with very unusual light curves, both of which
are rare.  Thus we are able to establish the CCSN rate by first counting
generic supernova light curves, then subtracting the relatively easier
to identify SNIa from the sample using Dilday's methods.  The
possibility of exotic supernova types that are
neither CCSN nor SNIa does exist, but their numbers as a fraction of
all supernovae detected to date is unlikely to
be an important consideration.  Uncertainty in the CCSN rate
due to exotic supernovae, using
our best estimate at their presence in our
sample, is negligible compared to other uncertainties.

The SDSS-III BOSS described in \citet{eisenstein,bossover} project provides
important ancillary data for
this result.  While SNIa not observed spectroscopically can have an 
estimate of their redshift based on their peak brightness and colors,
this is not possible for the much more diverse CCSN.  Unknown redshifts
for supernova candidates observed only photometrically would be a dominant
uncertainty for the CCSN rate, while limiting CCSN candidates to those that
were observed spectroscopically would limit the size of the sample 
leading to a large statistical uncertainty on the CCSN rate.  
The SDSS-III BOSS project provided a way out of this dilemma by
spectroscopically measuring the redshift of almost all $r\lesssim 21$ galaxies 
that are hosts to SDSS-II SNS supernova candidates.

At the beginning of the SDSS-III BOSS project, we compiled a list of suggested targets. 
This was a complete list of galaxies, not observed spectroscopically in the SDSS-II SNS, 
that are nearest to a supernova candidate in angular distance and nearest in isophotal 
distance.  Isophotal distance measures
the candidate's distance from the galaxy center, as a fraction of the
galaxy's isophotal size along the galaxy-candidate axis.  When a galaxy 
is nearest to a candidate in both angular and isophotal distance at a
redshift smaller than 0.1, confidence is high, better than 97\%,
that it is the galaxy in which the supernova candidate actually occurred.
For more details on the selection of the galaxies targeted
by SDSS-III BOSS see \citet{olmstead2013,campbell2013,sako2014}.

A small fraction of the recommended targets could not be observed due to technical 
reasons, usually fiber collisions, and observational constraints, usually the
requested target's brightness was below the reliable measurement threshold.
SDSS-III BOSS discovered that some, less than 5\%, of the targets
were variable stars or quasars, which unsurprisingly means that 
some of the variable
light sources observed by our survey were not a supernova at all.  For the
remainder, which appear to be typical galaxies, the BOSS team
measured redshifts.  The data were taken with the BOSS
spectrograph described in \citet{bossspec}
and processed by the BOSS pipeline described in \citet{bosspipeline}.

Our sample has a heterogeneous sample of spectra taken by many
different instruments of varied natures.  We make use of these spectra
to identify the supernova type and measure the redshift.  In the redshift
range of interest, less than 0.1, the supernova identification for SNIa
is very secure and the uncertainty on the redshift is negligible.

\section{Supernova Detection Efficiency Model}
\label{sec:sneff}

Our rate measurement begins with the complete set of $9933$ SDSS-II SNS
candidate light curves with associated redshifts smaller than 0.3 or
no good measure of their redshifts, as produced by the analysis pipeline outlined
in the preceding section.
The SDSS-II SNS used relatively loose
criteria for identifying a supernova candidate.  Therefore,
the candidate set from which we begin includes a large number of variable
objects that are not supernovae, such as variable stars, quasars and
other active galaxies, or perhaps even novae within the Milky Way.

To separate supernovae from other variable object types, we employed
the phenomenological light curve fitting method used by
the SNLS in \citet{bazin}.  They model observed
supernova brightness in each pass band as a function of time using the
formula:
\begin{equation}
\label{eqn:fbazin}
        f(t)  =  A
        \frac{e^{-(\frac{t-t_0}{\tau_{F}})}}
             {(1+e^{-(\frac{t-t_0}{\tau_{R}})})}
\end{equation}
We fit for the four parameters $A$,
the overall amplitude of the light curve, $t_0$, roughly
the time of peak luminosity, $\tau_R$, the rate of flux increase
long before the peak, and $\tau_F$, the rate of flux decrease long
after the peak.  We fit in the $r-$band and the maximum of the function
defines the peak brightness which we will use to model our detection efficiently.
Details of the fitting procedure are given in \citet{taylor}.

\begin{figure}[htbp]
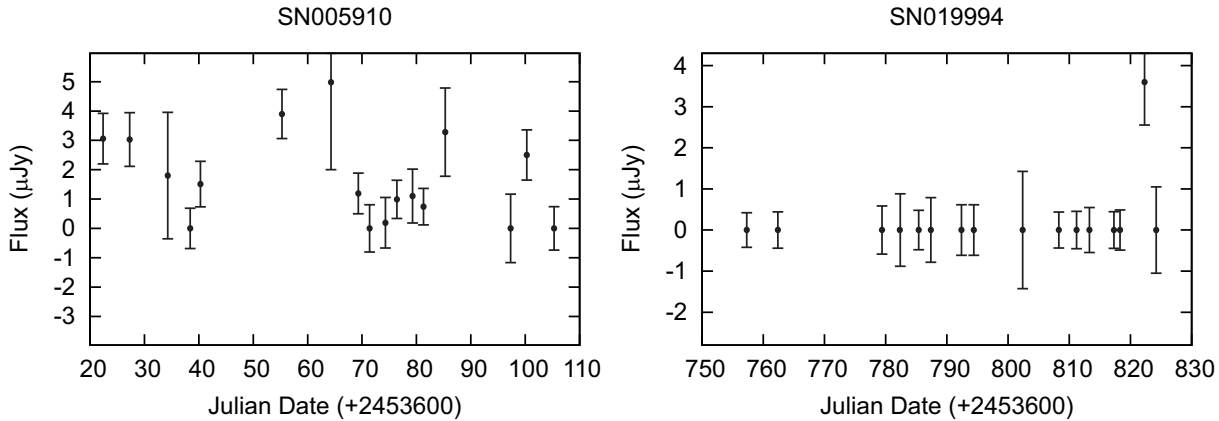

\begin{center}
\includegraphics[width=8.0cm]{SN005910.epsi}
\includegraphics[width=8.0cm]{SN019994.epsi}
\caption{SDSS-II SNS candidate $r-$band light curves are shown for two
objects which the light curve model did not converge to a best fit.
\label{fig:nofit}}
\end{center}
\end{figure}

Of 9933 light curves processed, the fit failed
to converge on 62 candidates.  Of those, 60 candidates recorded null or
negative flux in the $r-$band for all epochs; those 60 are
discarded.  These were identified by human scanners in the first
season based on an initial, later improved, photometric image subtraction.
Visual inspection confirms that the remaining two display
oscillatory features not characteristic of supernovae, as shown in
Figure~\ref{fig:nofit}.  
The uncertainty due to excluding non-converging fits is negligible compared
to other sources of uncertainty.

The supernova model adapted from \citet{bazin}
can be fit to virtually any light curve; however, the fit will be
poor for light curves without a clear, dominant peak.  To measure the
quality of the fit, we again follow the method of Bazin 
by fitting a second model to each light curve,
which is just the best-fit constant flux.

\begin{figure}[htbp]
\begin{center}
\includegraphics[width=8.25cm]{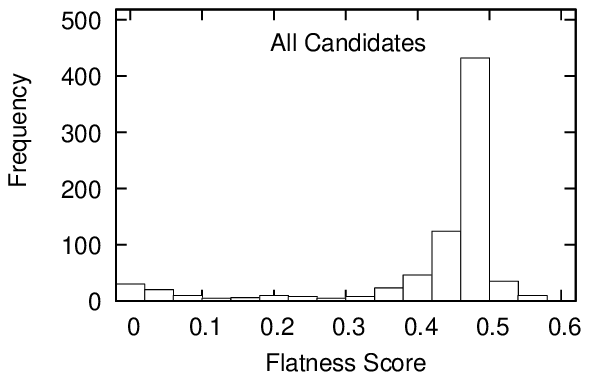}
\includegraphics[width=8.25cm]{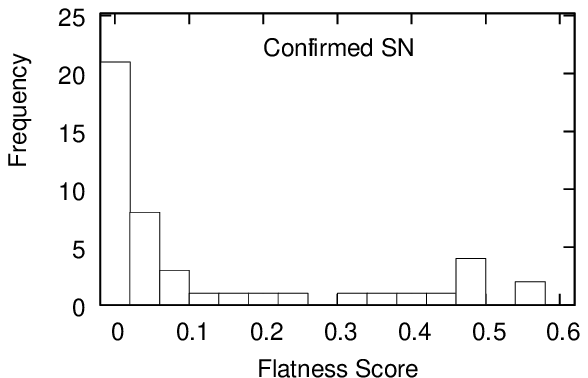}
\includegraphics[width=8.25cm]{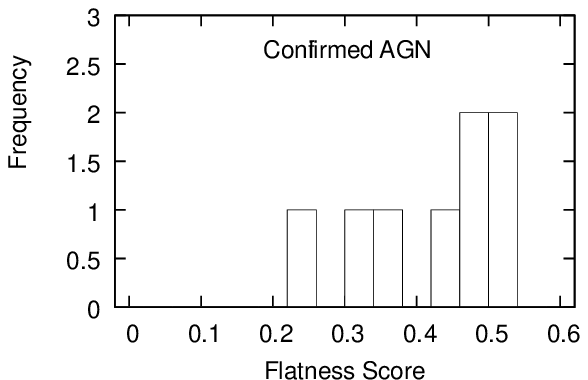}
\caption{Flatness score, as defined in the text, distribution is shown
for all candidates (top), for
spectroscopically confirmed supernovae (middle), and for confirmed AGN (bottom).
All data have $z < 0.09$.}
\label{fig:flatall}
\end{center}
\end{figure}

When the constant flux fit has a chi-squared comparable to the
chi-squared for the model fit, the object either is not a supernova, or
the data are too noisy to identify it as a supernova.  In either case,
we remove such objects from the rate sample.  The model and constant
flux chi-squared are directly compared, even though the model has
three more degrees of freedom, as both functions are fit
to the same number of data points, which in almost all cases is large
compared to the number of fit parameters.
To quantify this comparison, we assign
each light curve a flatness score, $\Lambda$, defined by:
\begin{equation}
\label{eqn:flatness}
	\Lambda \equiv
	\frac{\chi^2_{model}}{\chi^2_{model}+\chi^2_{const}}
\end{equation}
The value of $\Lambda$ ranges from zero for the best measured, obvious
supernovae, to one for light curves that show no supernova features.
Figure~\ref{fig:flatall} shows the distribution of $\Lambda$ for all
SDSS-II SNS candidates with redshift less than 0.09.  
It is bimodal, with a large peak near $\Lambda
= 0.5$, and a smaller peak near $\Lambda = 0$.

\begin{figure}[htbp]
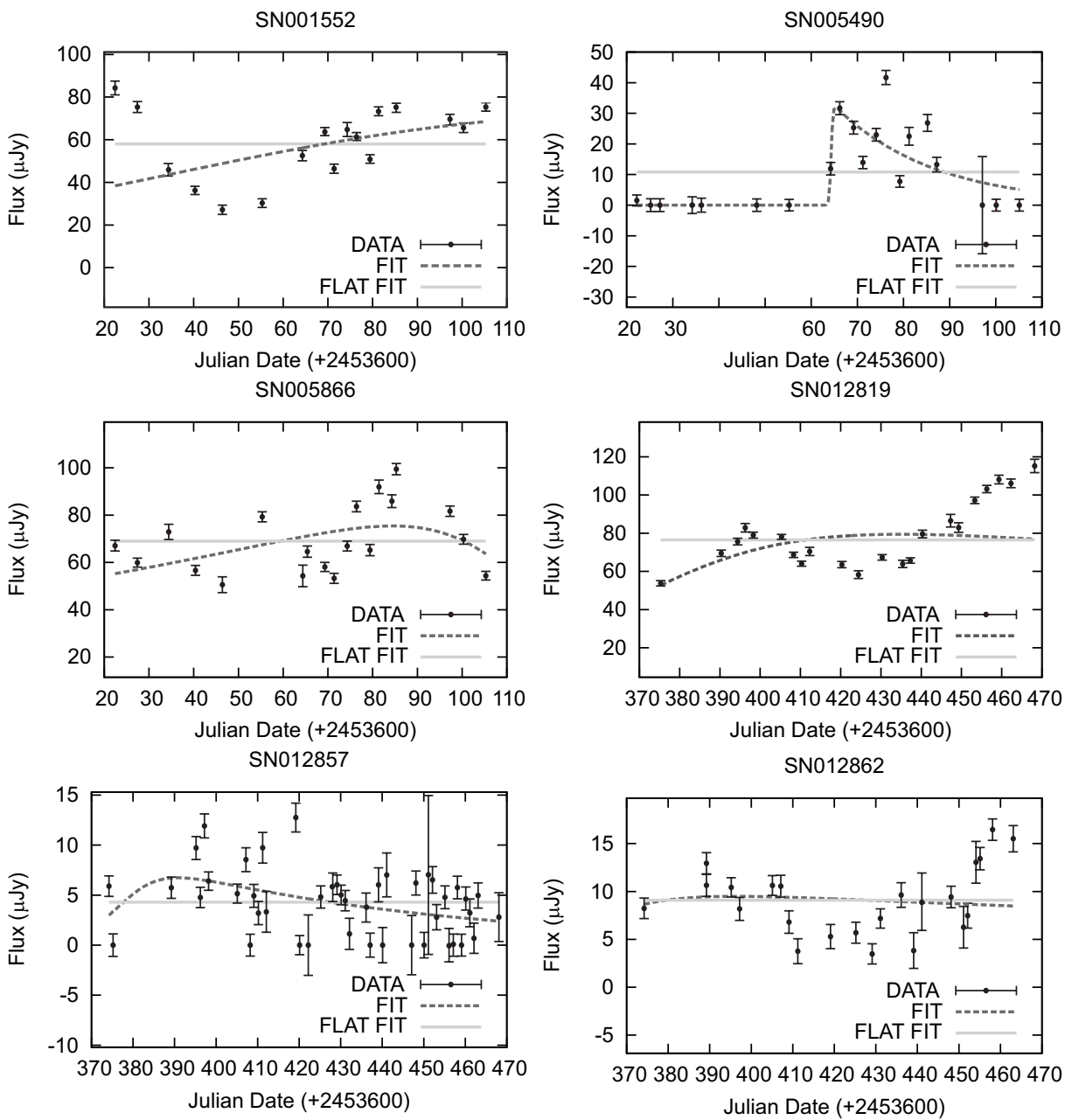

\begin{center}
\includegraphics[width=8.0cm]{SN001552.epsi}
\includegraphics[width=8.0cm]{SN005490.epsi}
\includegraphics[width=8.0cm]{SN005866.epsi}
\includegraphics[width=8.0cm]{SN012819.epsi}
\includegraphics[width=8.0cm]{SN012857.epsi}
\includegraphics[width=8.0cm]{SN012862.epsi}
\caption{Above are examples of confirmed AGN light curves at various redshifts, with the
the constant flux fit with a full line and the supernova
model described in the text with a dashed line. All these candidates were excluded by the flatness requirement.
Only the r-band is shown for clarity.
\label{fig:flatsample1}}
\end{center}
\end{figure}

\begin{figure}[htbp]
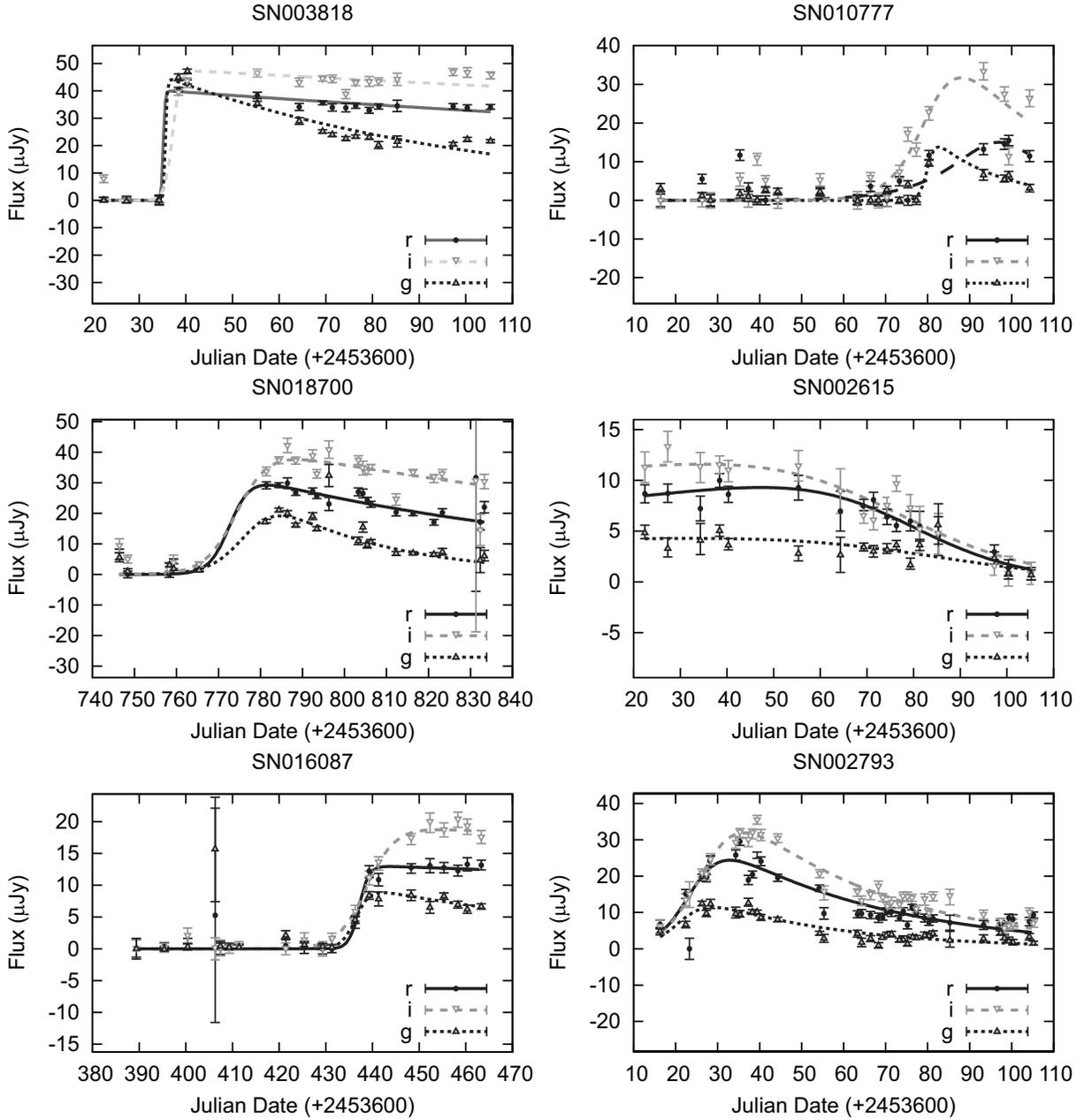

\begin{center}
\includegraphics[width=8.0cm]{SN003818.epsi}
\includegraphics[width=8.0cm]{SN010777.epsi}
\includegraphics[width=8.0cm]{SN018700.epsi}
\includegraphics[width=8.0cm]{SN002615.epsi}
\includegraphics[width=8.0cm]{SN016087.epsi}
\includegraphics[width=8.0cm]{SN002793.epsi}
\caption{Above are examples light curves accepted into the core collapse supernova
rate sample.  The lines show the SN model fit described in the text.
\label{fig:oksample1}}
\end{center}
\end{figure}

To test the correlation between flatness score and object type, we examine
the $\Lambda$ distribution for two candidate sub-samples:  core collapse
supernovae and active galaxies, both of which have been confirmed through
spectroscopic analysis, as described in \citet{sako1,sako2}.
We find that spectroscopically confirmed CCSN, the middle plot in Figure~\ref{fig:flatall},
are concentrated near $\Lambda = 0$, as expected, while the bottom plot shows that
confirmed AGN are concentrated near $\Lambda = 0.5$.
Based on these distributions we define supernova candidates to
have $\Lambda_c < 0.354$.  The systematic uncertainty of this choice
is discussed below.
A representative selection of light curves, confirmed AGN at
various redshifts, excluded by
the flatness score requirement
are shown in Figure~\ref{fig:flatsample1}, and
Figure~\ref{fig:oksample1} shows examples of accepted light curves.

The magnitude limit, beyond which a supernova is too dim for the survey
to detect, is not a sharp boundary.  There is a range of magnitudes,
due to light curve shape, observing conditions, and survey cadence, over
which supernovae have a finite probability of detection.  We refer to this
function as
our survey's detection efficiency.  
We model the detection
efficiency as a function of the supernova's peak apparent magnitude
as seen in the SDSS $r-$band.	The form of the model efficiency
function, $\epsilon(m)$ is:
\begin{equation}
\label{eqn:effic}
	\epsilon(m) = \frac{1}{2} \left[1-\mathrm{erf}(\frac{m-m_E}{\sigma_E})\right]
\end{equation}
This form is simply the convolution of a normal distribution of 
width $\sigma_E$ with a step function whose transition occurs at $m_E$.

\begin{figure}[htbp]
\begin{center}
\includegraphics[width=16.5cm]{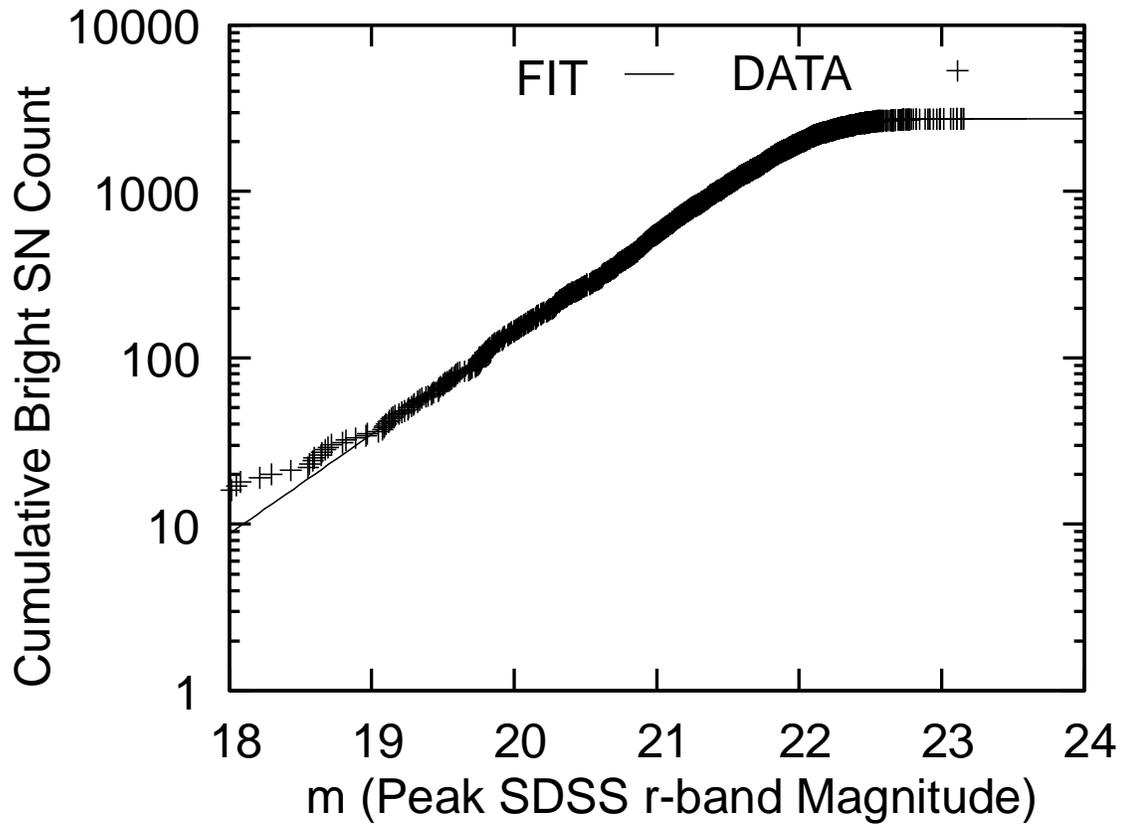}
\caption{The cumulative bright supernova candidate count, $N_p(m)$,  versus the 
peak $r-$band magnitude ($m$).  The fit shown with a line and discussed
in the text determines the efficiency model.
\label{fig:nm}}
\end{center}
\end{figure}

To measure $m_E$ and $\sigma_E$
we start with the sample of SN candidates that have peak
magnitudes bright enough that our detection efficiency is nearly one.
We take this magnitude to be 21; see below for further discussion.
The cumulative number of candidates with peak magnitude $m$ smaller than $m_E$,
$N_p(m)$, should have the form:
\begin{equation}
\label{eqn:magpop}
	N_p(m) = p_0 + p_1 10^{0.6m} + p_2 10^{0.8m} + p_3 10^{1.0m}
	+ p_4 10^{1.2m} 
\end{equation}
where the coefficients $p_i$ are determined by fitting the observed
distribution shown in Figure~\ref{fig:nm}.
Equation~\ref{eqn:magpop} is derived by first expressing the
SN rate, $\rho_{SN}$, as a function of luminosity distance, $D$,
and expanding $\rho_{SN}(D)$ with a Taylor series in $D$.
The luminosity distance is then in turn expressed in terms of the distance
modulus, $\mu$, according to:
\begin{equation}
\label{eqn:distmu}
	D = (10 \mathrm{pc}) \mbox{ } 10^{0.2\mu}
\end{equation}
Integrating over the sample volume and simplifying then yields 
Equation~\ref{eqn:magpop}, approximated to fourth order in $(10\ \mathrm{pc})/D$.

Once the best fit $N_p(m)$ is obtained for the bright portion of the
rate sample, it is compared to the actual number of candidates at
all magnitudes, $N_a(m)$.  The form of $N_p(m)$ should
be nearly the same at dimmer magnitudes as at bright magnitudes, since
each cohort in apparent magnitude includes objects at a wide range of
distances.  

\begin{figure}[htbp]
\begin{center}
\includegraphics[width=16.5cm]{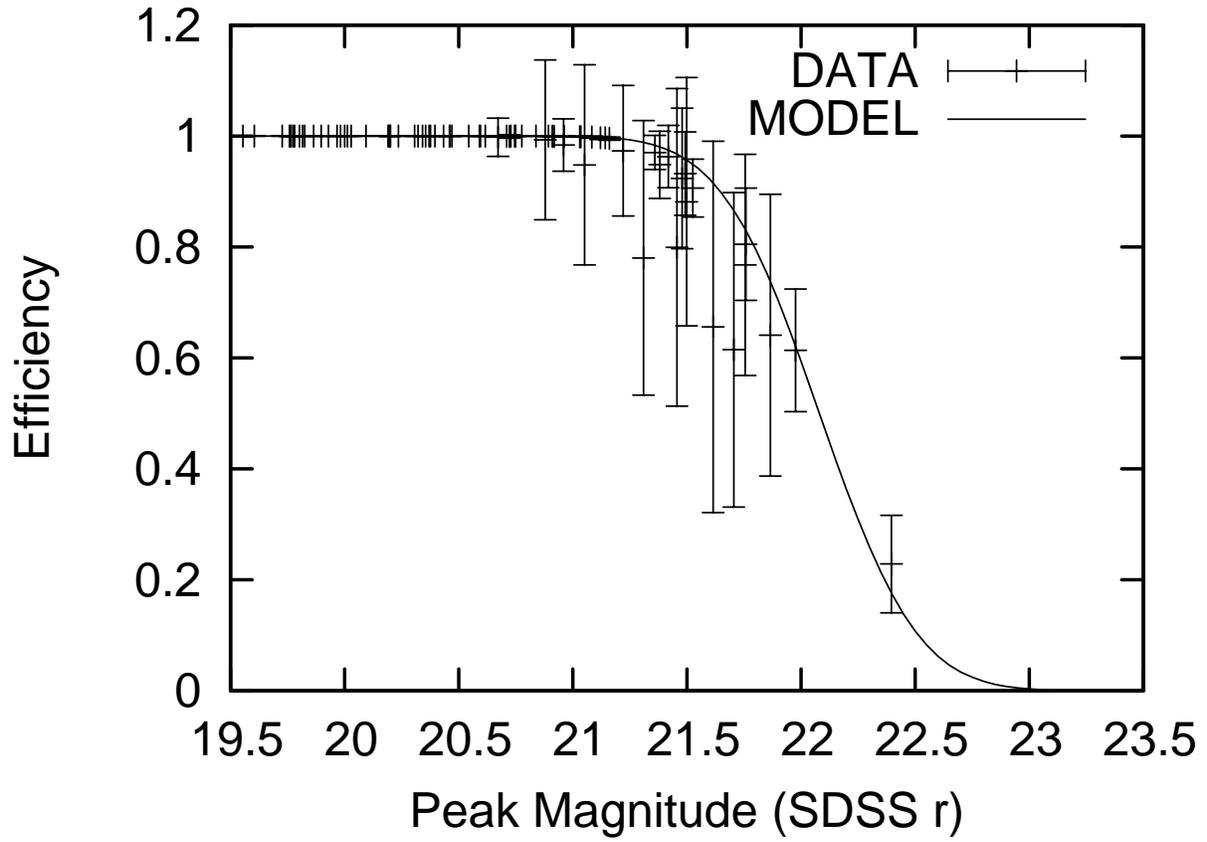}
\caption{The model supernova detection efficiency as a function
of the peak brightness for all SN candidates.  The model fit
shown with a line is used to correct the observed count of CCSN.
\label{fig:effic}}
\end{center}
\end{figure}

Next we fit the actual number of candidates, $N_a(m)$, to 
\begin{equation}
\label{eqn:efficfit}
	N_a(m) = \int_{-\infty}^m \! \epsilon(a) N_p(a)\,
	\mathrm{d}a
\end{equation}
which includes the effect of the efficiency.
The parameters of $N_p(m)$, the model population function, are
held fixed at the values fitted to the bright sub-sample, while the
parameters of $\epsilon(m)$, the efficiency function, are allowed
to float.  The resulting, best fit efficiency model is displayed in
Figure~\ref{fig:effic}.
Details for the efficiency model can be found in \citet{taylor}.

With the detection efficiency model in hand, we weight each supernova
candidate according to the inverse of the efficiency based on its peak
apparent magnitude.  The corrected supernova count is then the sum
of these weights, for all accepted candidates.	
In addition, we find the uncertainty in each candidates' weight by taking the
standard deviation of 1000 random trials.  For each trial the candidate's peak 
magnitude is drawn from a Gaussian distribution of width given by the uncertainty in the
candidate's peak magnitude weighted by the detection efficiency.

\begin{figure}[htbp]
\begin{center}
\includegraphics[width=16.5cm]{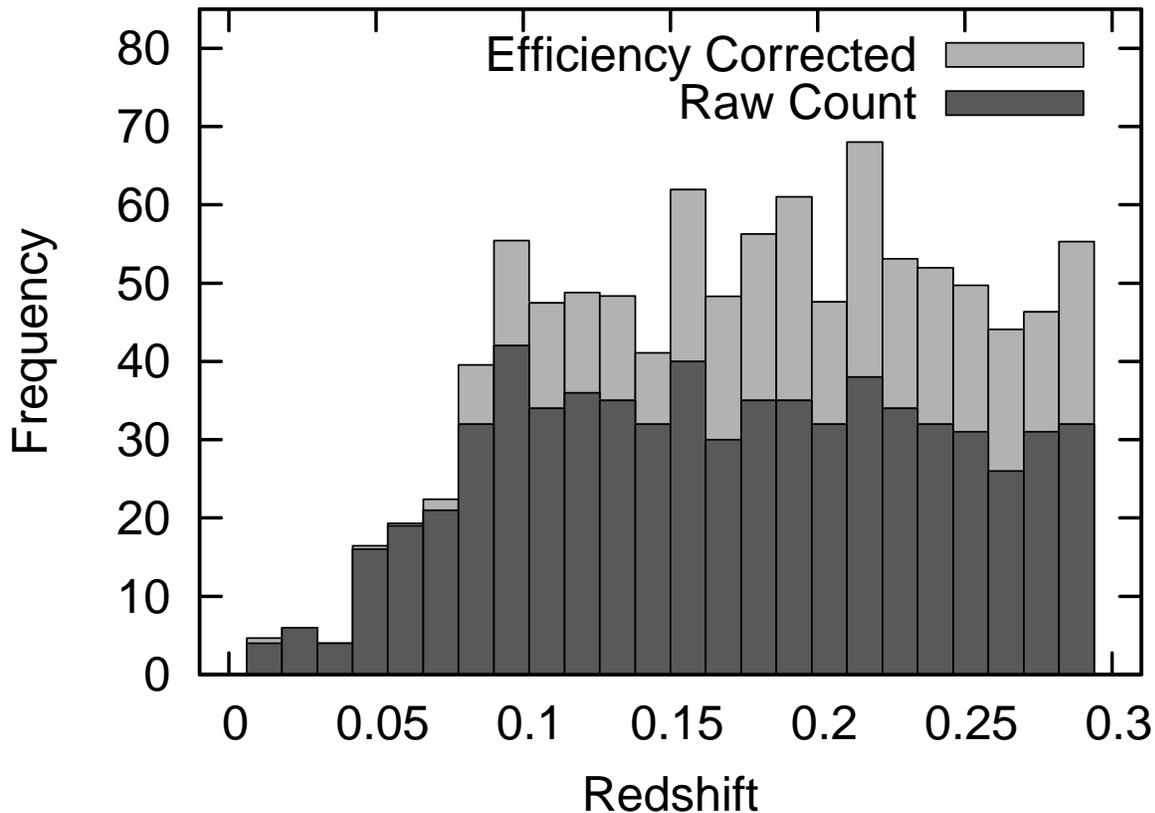}
\caption{The redshift distribution of the CCSN rate sample is shown,
sub-luminous sample excluded, before and after the efficiency
correction.
\label{fig:zehist}}
\end{center}
\end{figure}

The resulting, corrected redshift distribution is shown in
Figure~\ref{fig:zehist}.  Also displayed in the figure
is the uncorrected, raw distribution.  The actual supernova count is expected to
increase as approximately the third power of redshift, because of the increasing
volume sampled.  However, beyond $z \approx 0.1$, an increasingly large
fraction of the CCSN population is at magnitudes where the survey has
very low or zero detection efficiency.

\begin{figure}[htbp]
\begin{center}
\includegraphics[width=16.5cm]{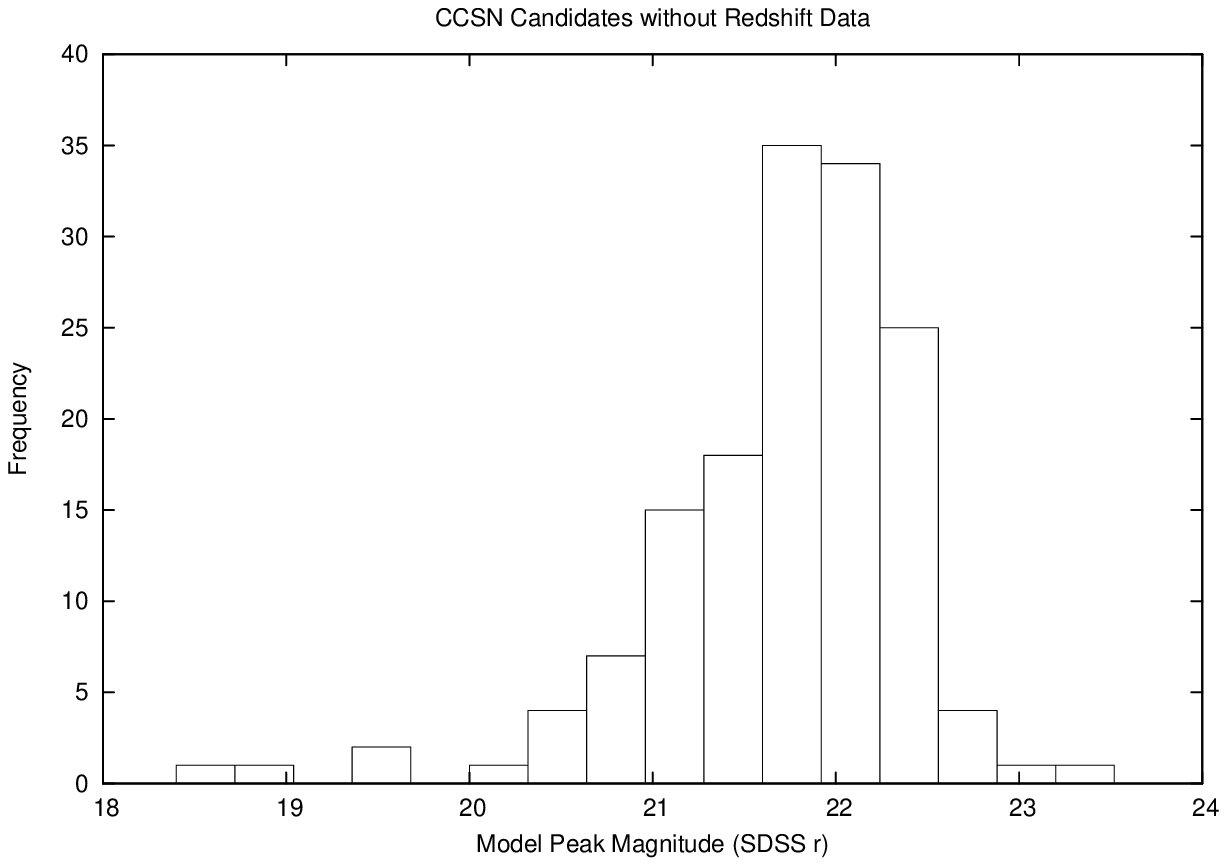}
\includegraphics[width=16.5cm]{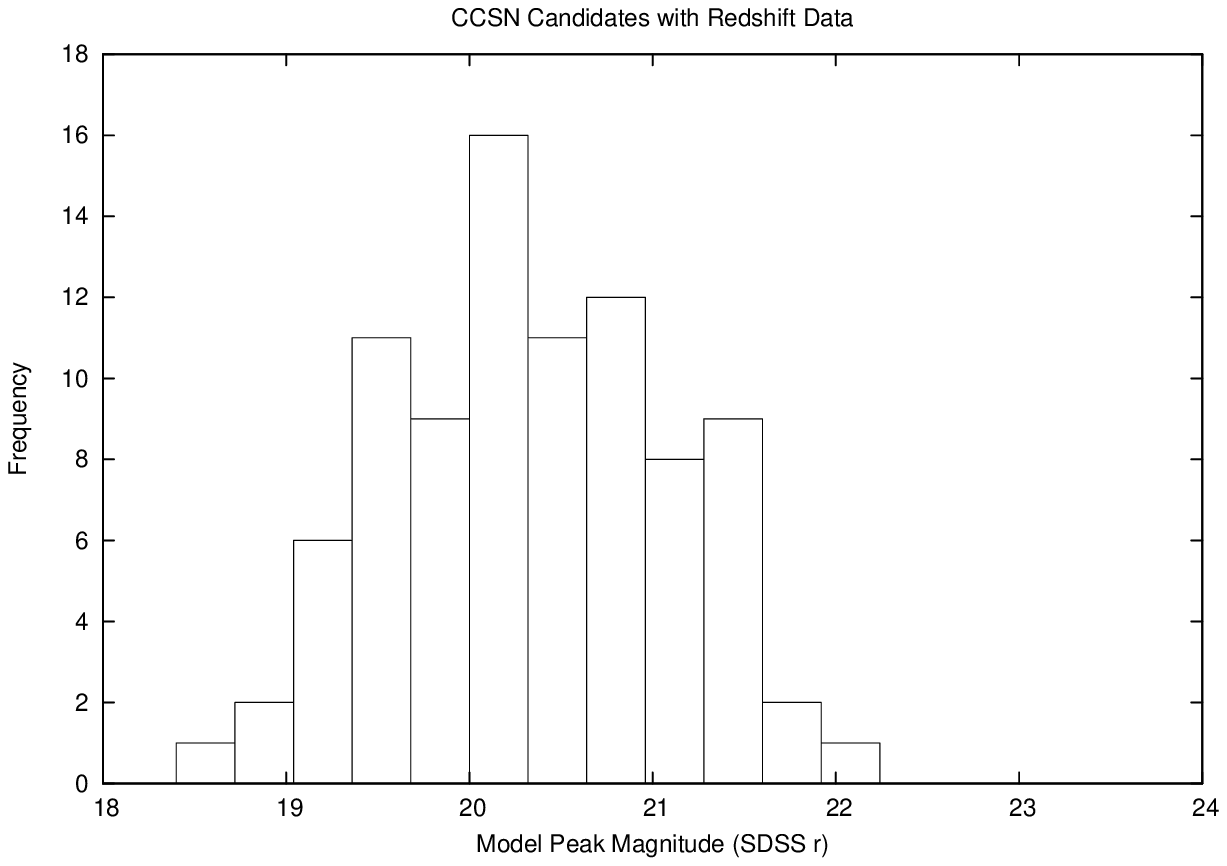}
\caption{The peak apparent magnitude distribution, in SDSS $r-$band,
is shown for SN candidates excluded from the rate sample for lack of
redshift information on top, and below for the SN cadidates in the rate
sample with redshift information.
\label{fig:noz}}
\end{center}
\end{figure}

We need a redshift as a distance measure.
Spectroscopic redshift of the supernova itself is preferred, but when
that is not available we instead use the spectroscopic redshift of the
host galaxy, or a photometric redshift estimate of the host galaxy as
a last resort.	However, in a few cases no spectra were taken of the
supernova, the host galaxy is too faint to detect, or for technical
reasons the redshift of the host galaxy is not measured, thus we have no
guidance at all for the candidate's distance.
We remove candidates from the sample due to lack of redshift
information, treating all of them as non-detections.  Figure~\ref{fig:noz}
shows the peak magnitude distribution of such candidates, showing
that the majority of these are at the edge of our detection limit.  
Also the figure shows the peak magnitude distribution of all the candidates
that pass all the other slections, but have a redshift determination.
By treating
candidates with no redshift as non-detections, we are implicitly assuming
that they are beyond our upper redshift cut of 0.09, and we attempt
to justify this assumption below.

\section{CCSN Rate Selections}
\label{sec:ccsncalc}

We select only candidates which fall in the Right Ascension (RA) and Declination (DEC)
for which the SDSS-II SNS had full coverage during its three seasons.  The RA had
a ragged edge night-to-night due to time and weather limitations and we simply chose the range
that was covered on 100\% of the observing nights.

The candidate light curve has
to be successfully fit by our light curve model as described above.
Candidates have to pass the flatness score test described above.
We truncate the rate sample
at $z = 0.09$ to limit the effect of low detection efficiency discussed above.
Galaxies with $z < 0.03$ are not homogeneously distributed, 
leading to a distortion in the apparent spatial distribution of
supernovae.  To avoid this effect, which violates assumptions of
our efficiency model, we exclude any supernova
candidate with redshift less than 0.03 from the rate sample.

We need to remove SNIa for our CCSN rate measure.
Fortunately, the SDSS-II SNS has sophisticated methods
for identification of SNIa, as they are the primary targets of interest
for the survey's cosmology mission.
Objects which display properties most similar to SNIa, and which are
suitable for spectroscopic observation, were submitted to the SDSS-II SNS
partner observatories.  The spectra obtained are subjected to
a cross-correlation analysis, comparing them with template spectra for
SNIa, CCSN and other object types.  The spectroscopic target selection
process and spectrum analysis are described in 
\citet{zheng}.
Those which match the SNIa templates are
marked as confirmed SNIa in the SDSS-II SNS database, and we remove all such
candidates from our CCSN rate sample.

The SDSS-II SNS identified many more supernova candidates than
available spectroscopic resources could observe, therefore methods were
developed to identify SNIa using only the photometric data from SDSS-II SNS
itself.	\citet{sako1,sako2} describes this process in detail;
we only summarize here.  Observed light
curves are fit 
to a variety of supernova templates, selecting the supernova type
which fits the data best.  Those candidates identified as SNIa by this
classification program are marked as photometric SNIa in the SDSS-II SNS
database, and we remove all such candidates from our CCSN rate sample.
Over 93\% of the SNIa candidates removed from our sample are
identified spetroscopically.

After fitting light curves to the \citet{bazin} model, 
we excluded all candidates
where the model peak occurs in the first or last 10 days of a observing
season, 2005, 2006, and 2007, or where there are no
observations before or after the time of the fit peak.  
The Julian date ranges included are shown in Table~\ref{tab:mjdrange},
summing to a total survey time of 264 days, or 0.723 years.
\begin{table}[htb]
\caption{Survey Time Ranges Included in CCSN Rate Measurement
\label{tab:mjdrange}}
\begin{center}
\begin{tabular}{lcr}
\hline
Observing Season & Julian Date Range & Days \\
\hline
2005 & 53622 - 53705 & 83 \\
2006 & 53974 - 54068 & 94 \\
2007 & 54346 - 54433 & 87 \\
\hline
Total & & 264 \\
\end{tabular}
\end{center}
\end{table}

As a supernova's
luminosity decreases, the effective volume of the survey
also decreases.  This means that the intrinsically dimmest
supernovae are always under-represented in any large volume of space.
We exclude from the rate sample any
supernova with peak absolute magnitude of $-15.0$ or dimmer in the SDSS
$r-$band assuming a flat spectral distribution.  We call these the
sub-luminous sample.
As all other supernova rate measurements to date suffer the same
limitation, either implicitly or via an explicit luminosity requirement such
as ours,
the results will still be comparable.
The final measurement
is therefore more properly the ``bright core collapse rate density.'' 
Note that here we are explicit as to what this means for this
result, while this is usually implicit in all other supernova
rate measurements.
We discuss the implications of this requirement further
in Section~\ref{sec:conc}.

Table~\ref{tab:cuts} summarizes the selections we have made on our initial large
sample to arrive at the 89 candidates we use for the CCSN rate density measurement.
\begin{table}[htb]
\caption{The SDSS-II SNS CCSN Rate Sample.
\label{tab:cuts}}
\begin{center}
\begin{tabular}{lr}
\hline
Selection                     & Number \\ \hline
Candidate Within RA-DEC Range & 9933 \\
Fit Light Curve               & 9871 \\
$0.03 < z < 0.09$             &  808 \\
Flatness Score Test           &  205 \\
Not Type Ia Supernova         &  157 \\
Within Observing Season       &  100 \\
Not Sub-Luminous              &   89 \\
\hline
Accepted CCSN Candidates  &   89 \\
\hline
\end{tabular}
\end{center}
\end{table}

\section{Corrections}
\label{sec:sncorr}

The SDSS-II SNS does not detect all supernovae that occur.
Weather, lunar phase, proximity to other celestial objects, and many
other factors can result in non-detection.  To compensate for this
effect, we constructed an efficiency model as described in Section~\ref{sec:sneff}.
Each candidate in the rate sample is then weighted by the inverse
of the survey detection efficiency, as a function of supernova peak
apparent magnitude.  The weighting scheme corrects the base sample count
upward by 7.44 supernovae.  The correction is relatively small as we
have confined our sample to a redshift range in which we are highly
efficient at detecting CCSN that are not sub-luminous.

The observed luminosity of a supernova may be less than its actual
luminosity due to extinction.
However, dust in each supernova's host galaxy, or perhaps even in the
local environment of the supernova itself, is more difficult to measure.
This local extinction is represented by the parametrization of 
\citet{cardelli},
which quantifies extinction 
with two parameters, $A_V$ and $R_V$. 

We employ the model developed by 
\citet{hatano}, 
who simulated an ensemble of host galaxies at random inclinations,
and placed model supernovae within them approximating the spatial
distribution of observed supernovae.  Dust along the resulting line of
sight determined the extinction of each model supernova.  The resulting
$A_V$ distribution is roughly exponential,
${P}(A_V) \approx \frac{1}{\tau_V} e^{-\frac{A_V}{\tau_V}}$,
with exponent $\tau_V = 0.50\pm 0.03$, where we use a standard $R_V = 3.1$.
Details of how we correct the observed supernova count due to
extinction can be found in \citet{taylor}.
The extinction
correction is small, less than 3\% of the measured rate, and
because of poorly
quantified uncertainties in the extinction model, we count 100\% of this
correction as a systematic uncertainty.  
There is evidence at higher redshifts that extinction for CCSN is
much larger \citet{melinder}, and this is another reason for
our redshift selection maximum at 0.09.

The results of efficiency and extinction corrections are summarized in
Table ~\ref{tab:corr}.
\begin{table}[htb]
\caption{Corrections to the CCSN Rate Sample Size.
\label{tab:corr}}
\begin{center}
\begin{tabular}{lrr}
\hline
Reason & Adjustment & Sample Size \\
\hline
Base Rate Sample & & 89.00 \\
Efficiency Correction & 7.44 & 96.44 \\
Extinction Correction & 2.11 & 98.55 \\
\hline
Corrected Rate Sample & & 98.55 \\
\end{tabular}
\end{center}
\end{table}

\section{Sources of Uncertainty}
\label{sec:uncert}

Table ~\ref{tab:err} summarizes the sources of statistical and systematic errors.
We discuss the systematic errors in more detail below.

To study systematic effects due to inherent assumptions in the efficiency model, we compared it to
another efficiency model.  This second model represented the detection efficiency as exactly 100\% 
for bright objects, zero for dim objects, and follows a cosine function of the object peak
 magnitude in the transition region between.  
The efficiency correction of 6.13 produced by this alternative model is 
consistent with the correction from our base model, 7.44, and the difference is small
compared to the statistical uncertainty on the base model which is calculated by summing
the uncertainty on the weight assigned to each candidate, $\pm7.43$. 
We conclude that the efficiency correction is not strongly dependent on the 
exact shape of the assumed detection efficiency 
function, and we assign the statistical error on the model we use, which
is large compared to the differences among these, as the systematic
uncertainty due to the efficiency correction.

We have cross checked this efficiency model with a SNANA,
a public supernova analysis and simulation package described in \citet{SNANA}, based
simulation of CCSN in the SDSS-II SNS.
It agrees that we are fully efficient for the bright CCSN, peak magnitude
brighter than 21, and rolls off smoothly to being completely inefficient for dimmer
CCSN, peak magnitude dimmer than 23.
We can obtain arbitrary precision with this simulation, but do not have a clear way to
assess its systematic uncertainty.  The simulation
is based on a finite set of CCSN templates
which are unlikely to span the diversity of real CSSN.  The uncertainty
from the empirically--derived efficiency described above is a better estimate
of our lack of understanding of our efficiency model than any procedure that we have
thought of based on a likely incomplete simulation.

The identification of supernovae based on flatness score, $\Lambda$,
also may introduce systematic uncertainty.  The number of confirmed AGN in the
target redshift region is small, and they may not be representative
of all AGN.  To estimate this uncertainty, we measure the rate at which
confirmed supernovae are rejected by the flatness score requirement, and
the rate at which confirmed AGN are accepted by the same requirement;
100\% of both numbers is taken as systematic uncertainty.  Adding these two
sources in quadrature, we find systematic uncertainty of $\pm 6.22$
in the supernova count due to misidentification.  This value is large compared
to our best guess of exotic supernovae types that could be contaminating our
sample.  \citet{li} notes that in a volume--limited sample
of supernovae lightcurves, 5\% of the SNIa sample is the distinct 
``2002cx-like'' objects, which we estimate as the number of truly
exotic objects that would not fall clearly under our two
categories of SNIa or CCSN.  This  result translates to $2.4$ objects
which we add
in quadrature to misidentification uncertainty to get $\pm 6.67$ as
an uncertainty on the supernova count.  
We get a similar estimate,
$\pm 3.5$ if rather we assume that truly exotic
objects are Type Iax supernovae, of which 2002cx is thought
to be a member, as suggested in \citet{iax}.
Type Iax are thought to be roughly 30\% of the SNIa sample in a given
volume, but their impact is reduced due to our brightness threshold
which about half of the dimmer Type Iax would not pass.
There is an even smaller change, $\pm 2$ if we
vary the flatness cut from $0.3$ to $0.4$ from its nominal value of 0.354.

At low redshifts, $(z < 0.1)$, where CCSN can
be detected with high efficiency, the combined photometric and spectroscopic
identification of SNIa is very accurate.  Our simulation tests in the 
SNIa rate measurement found that less than
1\% of SDSS-II SNS candidate SNIa are actually CCSN, as found in \citet{dilday2010}.
CCSN are more than three times as numerous as SNIa in a given volume
of space, therefore excluding survey-identified SNIa from our sample
should remove less than 0.3\% of CCSN from the rate calculation;
an insignificant effect compared to other sources of uncertainty.

Though the SDSS-III BOSS data greatly improved the accuracy of redshift
estimates, there are still some supernovae in the sample with only
photometric redshifts.	To estimate the resulting uncertainty, we compared
spectroscopic and photometric redshift measurements for those candidates
where both were available.  The resulting error is counted as the number
of candidates that would be moved into or out of the redshift range of
interest when replacing the photometric value with
the spectroscopic value.  This yields $\pm 10\%$ systematic uncertainty in
the supernova count, due to incorrect redshift.

For the good light curves without
any reasonable measure of their redshift we make an estimate
of the number that could violate our
assumption that they were beyond the redshift range of our rate sample.
We modeled each individual redshift as a Gaussian with mean and width
equal to the mean and width of the redshift distribution of the sample
all 1074 light curves that pass all our other selection,
have good redshifts, and peak magnitudes within one of the
candidate without redshift.  Many of these are SNIa at redshifts 
greater than 0.1.  We then integrated
the area of these 149 Guassians between 0.03 and 0.09 and find a sum of
less than one.  Our best guess is that a very small number of these, compared to the 
10\% count uncertainty due to incorrect redshifts,
could possibly enter our sample and we assign no systematic uncertainty
due to removing them.  Our assertion that the almost all of the
sample without good redshifts are beyond the redshift range of our rate sample.

The supernova rate density is measured by the supernova count divided by survey
time and volume.  The error on the time and volume is negligible compared to error
in the supernova count.  We use the co-moving
volume that encloses the sample of supernovae.	
Uncertainty in cosmological parameters could make a small contribution to
uncertainty in the co-moving volume, and hence uncertainty in the supernova
rate density.  To probe the effect of varying cosmological parameters on
co-moving volume, we employed the iCosmos cosmological distance calculator,
developed by \citet{vardanyan}.
The SDSS-II SNS cosmology study found a 5\% uncertainty
in $\Omega_m$ in \citet{kessler}, which translates to approximately 0.3\%
uncertainty in comoving volume at $z = 0.09$.  We ignore this.  
\citet{kessler} also found approximately 10\% uncertainty
in $w$, the dark energy equation of state parameter,
which translates to a 1.3\% uncertainty in comoving volume
at $z = 0.09$.	The effect of uncertainty in $w$ is small, but it is
significant enough to account for in sources of uncertainty for the CCSN rate,
and we have included it in the uncertainty tabulation.

Table ~\ref{tab:err} summarizes the sources of statistical and systematic
uncertainty; they are added in quadrature for a total uncertainty estimate.
\begin{table}[htb]
\caption{Sources of Uncertainty in the CCSN Rate
\label{tab:err}}
\begin{center}
\begin{tabular}{lrr}
\hline
Source of Uncertainty                & Uncertainty in SN Count & \% Uncertainty \\ \hline
Statistical Uncertainty              & 10.31 & 10.6\% \\ \hline
Efficiency Correction                &  7.43 &  7.5\% \\
Extinction Correction                &  2.11 &  2.1\% \\
Identification by Flatness           &  6.67 &  6.8\% \\
Redshift Uncertainty                 & 10.00 & 10.2\% \\
Volume due to Cosmology Parameters   &       &  1.3\% \\ \hline
All Systematic Uncertainties         &       & 14.6\% \\ \hline
Total Uncertainty                    &       & 18.0\% \\
\end{tabular}
\end{center}
\end{table}

\section{CCSN Rate Calculation}
\label{sec:rate}

The survey volume is calculated as the difference between two sections of
a sphere, each subtended by the solid angle given by the chosen limits
on right ascension and declination.  Because the declination range lies
within $1.25^{\circ}$ of the celestial equator, we can use the small
angle approximation for $\Delta\theta$ given by:
\begin{equation}
\label{eqn:volcalc}
	V  =  \frac{1}{3} ( \Delta\theta ) ( \Delta\phi ) ( D_{c2}^3 -
	D_{c1}^3 )
\end{equation}
where $D_{c1}$ and $D_{c2}$ are the co-moving distances at $z = 0.03$
and $z = 0.09$, respectively.  The right ascension range, which gives
$\Delta\phi$, of the survey is $-50^\circ$ to $50^\circ$.
Note that we further correct by a factor of $1/(1+z)$ in the survey
volume integration to account for the cosmological time dilation.
This calculation yields a survey volume
of $1.29 \times 10^6\ {\mathrm Mpc}^3$.  Uncertainty in cosmological parameters
introduces an error in the calculated volume, which are propagated
into the systematic error on the rate as discussed above.  The survey area 
excluded due to bright stars, known variables, etc. is less than 2\%.

When counting supernova that occurred during the survey time range, a
systematic error may be introduced by incorrectly measuring the peak time.
Light curves only have data recorded every two days, at best, therefore
the light curve fit may incorrectly judge the time of maximum luminosity.
To measure error due to this effect, we divided each observing season into
two halves; the difference in supernova count between the two halves was
5.5, with an average of 44.5 supernovae in each half.  This result is within the
bounds of statistical fluctuation, since $5.5 < \sqrt(44.5)$, therefore
we conclude that additional, systematic error to to time uncertainty
is negligible.  Our SNANA based simulation, where we can compare the
real peak time with an observed peak time, also finds that
the fluctuations in the number of accepted light curves due to
mismeasurements of the peak time are small compared the
statistical uncertainty.

The final rate calculation is just the supernova count, divided by
the survey time and volume.  A factor of $h^3$, with $h = 0.7$ is included in
the rate units, to reflect the fact that this rate density measurement scales with the
Hubble constant, $\mathrm{H}_0$, with a value of $\mathrm{H}_0 =
h\ 100 \mathrm{km/s} \mathrm{Mpc}^{-1}$.  The final rate is given by:
\begin{equation}
\label{eqn:finalrate}
\rho_{\mathrm{CCSN}} = \frac{98.55 \mathrm{CCSNe}}
                            {1.29 \times 10^6 \mathrm{Mpc}^3 \times 0.723 \mathrm{yr}}
	             = (1.06 \pm 0.11 (stat) \pm 0.15 (sys)) \times 10^{-4}
	\frac{(h/0.7)^3}{\mathrm{Mpc}^3 \mathrm{yr}}
\end{equation}

\section{Discussion}
\label{sec:conc}

We compute the mean of our sample's redshifts weighting by the inverse
of the efficiency correction to measure an average redshift for
our rate density of $0.072 \pm 0.005$.  A more complex procedure which
tries to take into account the redshift dependence of the rate described in 
\citet{taylor} finds an average redshift of $0.080 \pm 0.005$ and we take the difference
as a systematic uncertainty on the redshift of our rate to get a final
value of $0.072 \pm 0.009$.

\begin{figure}[htbp]
\begin{center}
\includegraphics[width=16.5cm]{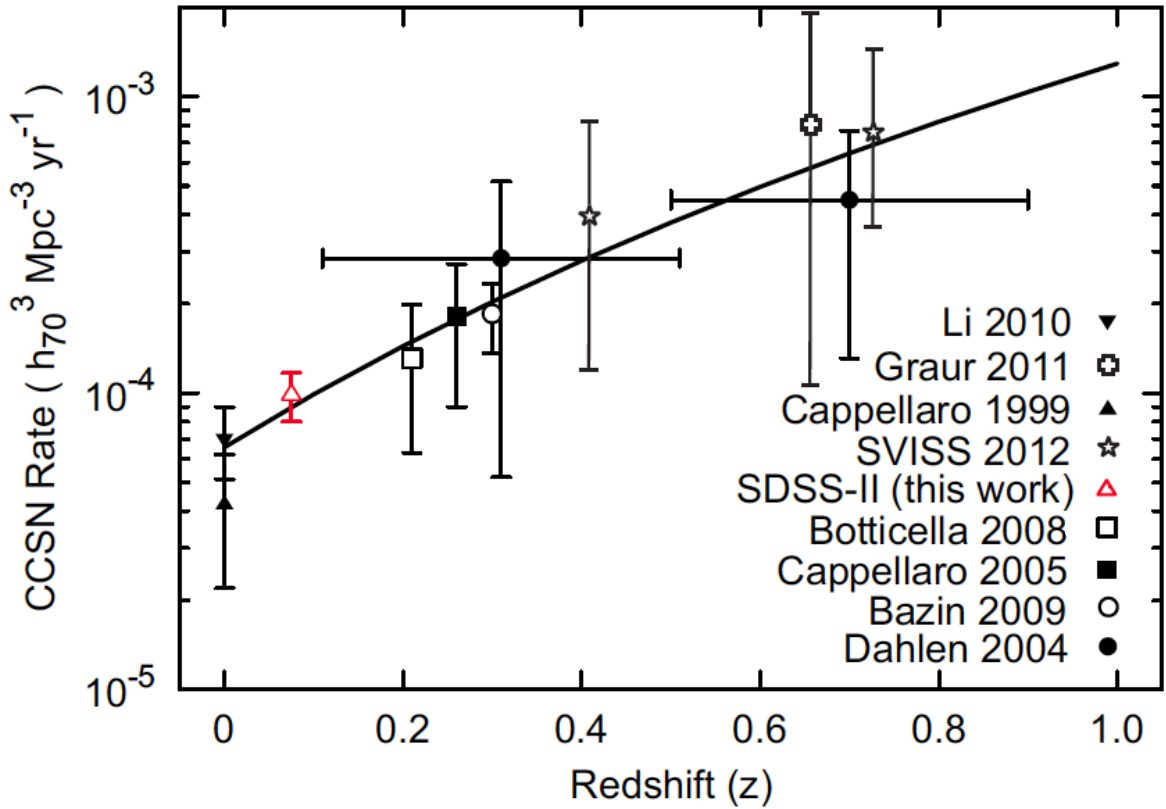}
\caption{The CCSN rate measurement from this work in red is shown with
previous CCSN rate measurements in the literature.
The solid line is the star formation rate, scaled to fit the
measured CCSN rates.
\label{fig:ratez}}
\end{center}
\end{figure}

Figure~\ref{fig:ratez} shows our result, and CCSN rate measurements
from the literature, versus the redshift.  The solid line
shows the expected trend in CCSN rate according to the increase in star
formation rate with redshift,
$\rho_{\mathrm{CCSN}} = \rho_0(1+z)^\beta$ with $\beta = 4.3$ as seen in
\citet{karim}, scaled to the 
combined rate measurements.
Our result is generally consistent with the trend identified in earlier
CCSN surveys, but provides coverage in a redshift range not previously
measured.  

We would like to compare the CCSN rate with the star formation rate.
Unfortunately establishing exact correspondence between these two rates
is complicated by our inability to observe
the dimmest CCSNe.  This includes both CCSNe that are intrinsically dim,
and those that are obscured by dust in the host galaxy.  In an effort to
better understand this sub-luminous CCSN population, we have attempted to
characterize the full distribution of CCSN absolute magnitudes, as much as
our data will allow.

\begin{figure}[htbp]
\begin{center}
\includegraphics[width=16.5cm]{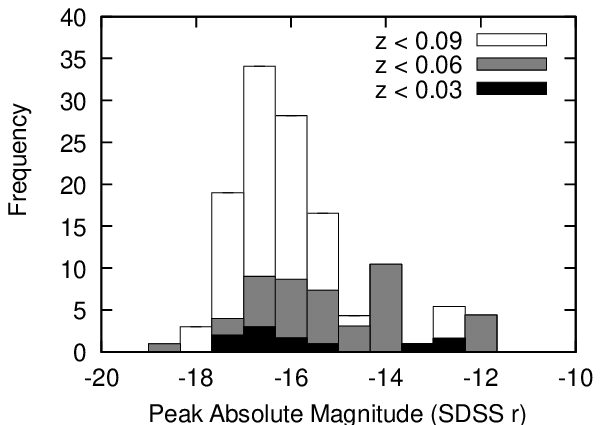}
\caption{The CCSN luminosity distribution, see text, that we derive for the rate sample in
this work, plus candidates excluded only because they were faint or below
the redshift range of the rate measure.
\label{fig:absmag}}
\end{center}
\end{figure}

Figure~\ref{fig:absmag} shows the efficiency-weighted absolute magnitude
distribution of our CCSN rate sample, plus those candidates excluded only
because they were too faint, dimmer than $M = -15$, or at redshift below our 0.03 limit.  
This is related to the core collapse supernova luminosity function, but
does not include corrections for the redshift dependence of our data.
The figure suggests that sub-luminous supernovae may be
underrepresented in the full rate sample, because a similar number
are detected in the small volume lowest redshift and the larger
volume mid-redshift bin.  To estimate the actual rate of
sub-luminous supernovae, we examine the sub-samples for $z < 0.03 $
and $0.03 < z < 0.06$.  The sub-luminous fractions in these ranges 
are 20\% and 24\%, respectively.  
Our sample is too small to regard these figures as
conclusive, but they do suggest that a sub-luminous fraction of 50\%,
the number required to solve rate problem in \citet{horiuchi}, 
is unlikely.  Our data do suggest that there might be a substantial
fraction, more than 20\%, of sub-luminous supernovae that have been missed
in existing rate measurements.  This is also suggested in \citet{mattila}.

Future surveys
promise orders of magnitude increase in the number of
supernovae observed.  While this
will greatly reduce statistical uncertainty, systematic uncertainty
may remain comparable to present day surveys, especially because only a
small fraction of such events are likely to have spectroscopic redshift measurements.
It is possible that almost all galaxy hosts can have
their redshifts measured by a massively parallel spectrometer yielding
a larger sample of CCSN with secure redshifts which will be important in better
measuring the rate of sub-luminous CCSN, a crucial factor in understanding
the correspondence between CCSN rate and star formation rate.
Also, the increased statistics will
be invaluable in more complex supernova measurements, such as understanding
the distribution of various supernova characteristics within the CCSN
population, and correlating CCSN rate density and characteristics with properties
of the host galaxy and location within the galaxy.

\section{Conclusion}
\label{sec:conc2}

In conclusion we have measured a bright core collapse supernovae rate density of
$(1.06 \pm 0.19 \times 10^{-4}) (h/0.7)^3/(\mathrm{Mpc}^3\ \mathrm{yr})$
at a mean redshift of $0.072 \pm 0.009$.  This agrees with the expectation
from previous measures and lies along the previously observed trend.  Our new
result is the most accurate single measurement and is at a newly explored 
region in redshift.  It derives
from a sample of 89 light curves observed with the SDSS-II SNS in the redshift
range of 0.03 to 0.09 that have been corrected with an efficiency derived
from our data and an extinction model from Hatano and collaborators in \citet{hatano}.

\acknowledgements

{\bf Acknowledgments}

The SDSS-II was managed by the Astrophysical Research Consortium 
for the Participating Institutions. The Participating Institutions were the 
American Museum of Natural History, Astrophysical Institute Potsdam, 
University of Basel, Cambridge University, Case Western Reserve University, 
University of Chicago, Drexel University, Fermilab, the Institute for 
Advanced Study, the Japan Participation Group, Johns Hopkins University, 
the Joint Institute for Nuclear Astrophysics, the Kavli Institute for 
Particle Astrophysics and Cosmology, the Korean Scientist Group, the 
Chinese Academy of Sciences (LAMOST), Los Alamos National Laboratory, 
the Max-Planck-Institute for Astronomy (MPA), the Max-Planck-Institute 
for Astrophysics (MPiA), New Mexico State University, Ohio State University, 
University of Pittsburgh, University of Portsmouth, Princeton University, 
the United States Naval Observatory, and the University of Washington.

This work is based in part on observations made at the following telescopes. 
The Hobby- Eberly Telescope (HET) is a joint project of the University of 
Texas at Austin, the Pennsylvania State University, Stanford University, 
Ludwig-Maximillians-Universit\"at M\"unchen, and 
Georg-August-Universit\"at G\"ottingen. The HET is named in honor of its 
principal benefactors, William P. Hobby and Robert E. Eberly. The Marcario 
Low-Resolution Spectrograph is named for Mike Marcario of High Lonesome 
Optics, who fabricated several optical elements for the instrument but 
died before its completion; it is a joint project of the Hobby-Eberly 
Telescope partnership and the Instituto de Astronom\'ia de la Universidad 
Nacional Aut\'onoma de M\'exico. The Apache Point Observatory 3.5 m 
telescope is owned and operated by the Astrophysical Research Consortium. 
We thank the observatory director, Suzanne Hawley, and former site manager, 
Bruce Gillespie, for their support of this project. The Subaru Telescope 
is operated by the National Astronomical Observatory of Japan. The William 
Herschel Telescope (WHT) is operated by the Isaac Newton Group, the Nordic 
Optical Telescope (NOT) is operated jointly by Denmark, Finland, Iceland, 
Norway, and Sweden, and the Telescopio Nazionale Galileo (TNG) is operated 
by the Fundaci\'on Galileo Galilei of the Italian INAF (Istituto Nazionale 
di Astrofisica) all on the island of La Palma in the Spanish Observatorio 
del Roque de los Muchachos of the Instituto de Astrof\'isica de Canarias. 
Observations at the ESO New Technology Telescope at La Silla Observatory 
were made under programme IDs 77.A-0437, 78.A-0325, and 79.A-0715. Kitt 
Peak National Observatory, National Optical Astronomy Observatories (NOAO), 
is operated by the Association of Universities for Research in Astronomy, 
Inc. (AURA) under cooperative agreement with the NSF. The South African 
Large Telescope (SALT) of the South African Astronomical Observatory is 
operated by a partnership between the National Research Foundation of 
South Africa, Nicolaus Copernicus Astronomical Center of the Polish 
Academy of Sciences, the Hobby-Eberly Telescope Board, Rutgers University, 
Georg- August-Universit\"at G\"ottingen, University of Wisconsin-Madison, 
University of Canterbury, University of North Carolina-Chapel Hill, 
Dartmouth College, Carnegie Mellon University, and the United Kingdom 
SALT consortium. The WIYN Observatory is a joint facility of the 
University of Wisconsin- Madison, Indiana University, Yale University, 
and NOAO.  The W.M. Keck Observatory is operated as a scientific partnership 
among the California Institute of Technology, the University of California, 
and the National Aeronautics and Space Administration. The Observatory was 
made possible by the generous financial support of the W. M. Keck Foundation.

This work was supported in part by the U.S. Department of Energy under 
contract number DE-AC0276SF00515 and the National Science Foundation.

Funding for SDSS-III has been provided by the Alfred P. Sloan
Foundation, the Participating Institutions, the National Science
Foundation, and the U.S. Department of Energy Office of Science.
The SDSS-III web site is {\url http://www.sdss3.org/.}

SDSS-III is managed by the Astrophysical Research Consortium for the
Participating Institutions of the SDSS-III Collaboration including the
University of Arizona,
the Brazilian Participation Group,
Brookhaven National Laboratory,
University of Cambridge,
Carnegie Mellon University,
University of Florida,
the French Participation Group,
the German Participation Group,
Harvard University,
the Instituto de Astrofisica de Canarias,
the Michigan State/Notre Dame/JINA Participation Group,
Johns Hopkins University,
Lawrence Berkeley National Laboratory,
Max Planck Institute for Astrophysics,
Max Planck Institute for Extraterrestrial Physics,
New Mexico State University,
New York University,
Ohio State University,
Pennsylvania State University,
University of Portsmouth,
Princeton University,
the Spanish Participation Group,
University of Tokyo,
University of Utah,
Vanderbilt University,
University of Virginia,
University of Washington,
and Yale University.



\bibliographystyle{apj}

\end{document}